\documentclass[useAMS,usenatbib,usegraphicx]{mn2e}
\usepackage{epsf}
\title[Bridging Effect of Void Filaments]
{The Bridge Effect of Void Filaments}
\author[Daeseong Park and Jounghun Lee]
{Daeseong Park\thanks{E-mail: pds2001@astro.snu.ac.kr} and
Jounghun Lee\thanks{E-mail: jounghun@astro.snu.ac.kr}\\
Department of Physics and Astronomy, FPRD, Seoul National University,
Seoul 151-747, Korea \\}

\begin{document}

\date{Accepted 2009 ???. Received 2009 ???; in original form 2009 May 27}

\pagerange{\pageref{firstpage}--\pageref{lastpage}} \pubyear{2008}

 \maketitle

\label{firstpage}

\begin{abstract}
Cosmic filaments play a role of bridges along which matter and gas
accrete onto galaxies to trigger star formation and feed central
black holes. Here we explore the correlations between the intrinsic
properties of void galaxies and the linearity $R_L$ of void
filaments (degree of filament's straightness). We focus on void
regions since the bridge effect of filaments should be most
conspicuous in the pristine underdense regions like voids. Analyzing
the Millennium-Run semi-analytic galaxy catalogue, we identify void
filaments consisting of more than four galaxies (three edges) and
calculate the means of central black hole mass, star formation rate,
and stellar mass as a function of $R_L$. It is shown that the void
galaxies constituting more straight filaments tend to have higher
luminosity, more massive central black holes and higher star
formation rate. Among the three properties, the central black hole
mass is most strongly correlated with $R_L$. It is also shown that
the dark halos constituting straight filaments tend to have similar
masses. Our results suggest that the fuel-supply for central black
holes and star formation of void galaxies occurs most efficiently
along straight void filaments whose potential wells are generated by
similar-mass dark halos.
\end{abstract}


\section{INTRODUCTION}\label{intro}

The spatial distribution of galaxies in the universe is quite
anisotropic, forming a filamentary web on large scales, which is
often called the {\it cosmic web}. The existence of cosmic web has
been explained as a natural phenomenon caused by the large-scale
coherence in the primordial tidal field \citep{bon-etal96}. The
physical properties of the galaxies located in the cosmic web are
believed to be related to the merging/accretion history of the
underlying dark halos, which in turn are affected not only by the
local density but also by the tidal fields as manifested by the
cosmic web. Although the effect of the tidal field on the galaxy's
properties is expected to be weaker than that of the density field,
several numerical and observational evidences have recently been
reported for the tidal influences on dark halos and galaxies. For
instance, \citet{des08} has shown by N-body simulations that the
tidal effect from the surrounding matter distribution causes earlier
virializations of dark halos embedded in overdense environments.
\citet{hah-etal08} have claimed using the results from
high-resolution N-body experiments that the high densities of the
regions with strong tidal forces induce the assembly bias
\citep{gao-etal05}. The link between the tidal field and galaxy
properties have been also detected in real observations. It has been
recently shown that the galaxy morphology and luminosity depend on
the large scale tidal field \citep{lee-erd07,lee-lee08}. When the
density is constrained to a narrow range, more luminous and
late-type galaxies are found to reside in regions with higher
ellipticity.

In previous works, however, what has drawn little attention is the
effect of the directions of the tidal forces on the galaxy
properties. If the halo formation and the merging/accretion events
occur preferentially along filaments, i.e., in the directions of the
principal axes of the tidal field, the tendency of the spatial
alignments of the tidal field should be taken into account when the
tidal influences on the galaxy properties are explored. Very
recently, \citet{por-etal08} have reported that the star formation
rate of a galaxy is observed to be enhanced when it falls along a
supercluster filament. This observational evidence suggests that
the merging and accreting along the coherent directions of the tidal
field trigger the star formation more efficiently.

In practice, the degree of the tidal alignment can be quantified by measuring
the straightness of a filament. The more straight a filament is, the stronger
the tidal alignment is.  Here, we propose a hypothesis that the highly
coherent accretion of matter and flow of gas along straight filaments
trigger the star formation and feed the central black holes most efficiently.
We call it the {\it bridge effect} of the filaments. To test this hypothesis,
we investigate the cross-correlations between the stellar mass, central black
hole mass, and star formation rate of the void galaxies and the straightness
of the filaments where the galaxies are located.  The reason that we consider
only void regions for this investigation is that the strong effect of the
density field can be controlled to a minimum level in the extreme underdense
regions like voids.

The organization of this paper is as follows. In \S 2, we analyze
the data from the Millennium Run semi-analytic catalogue. In \S 3,
we explore the correlations between the properties of void galaxies
and the linearity of host filaments. A summary of the results and
the final conclusion are given in \S 4.

\section{DATA}

We use the void catalogue that was constructed in our previous work
\citep{par-lee07a} by applying the void-finding algorithm of
\citet[][hereafter HV02]{hoy-vog02} to the Millennium-Run semi-analytic
galaxy catalogue \citep{spr-etal05}.
As done in our companion paper \citep[][hereafter PL09]{par-lee09}, we first
select those large voids which contain more than $30$ galaxies and then apply
the filament finding algorithm of \citet{bar-etal85} to the selected voids.
The filament finding algorithm of \citet{bar-etal85} basically utilizes the
minimal spanning tree (MST) technique to identify filamentary structures.
A detailed description of the filament-identification in the Millennium
voids is provided in PL09. Here we provide a concise summary of it.

For each selected void, we construct a MST of the void galaxies as
follows: A starting galaxy (i.e., a node) is linked to its nearest
neighbor galaxy by a straight line (i.e., an edge). This partial
tree is extended by connecting it to the next nearest neighbor void
galaxy. When all galaxies in a given void are connected, a MST is
obtained. Then, we reduce the MST of each void to identify the
filaments through pruning and separating processes. Each MST is
pruned at the $p$-level by removing the small-scale twigs. The
pruned MSTs are separated into several distinct shorter pieces by
removing those edges longer than a given length-threshold, $l_c$.
The level of pruning and the length threshold are set at $p=5$ and 
$l_{c}=\bar{l} +\sigma_{l}$, respectively, where $\bar{l}$ and $\sigma_{l}$ 
denote the mean edge-length averaged over all unreduced MSTs and its 
standard deviation, respectively (PL09). It is worth mentioning 
that the pruning level was set at $p=4$ in PL09 where the void filaments 
were found from the dark halos identified by the Friends-of-Friends (FOF) 
algorithm since the size distribution of void filaments were found to 
become stabilized at $p\ge 4$. Now that the void filaments are found 
using not the FOF halos but the galaxies in the current analysis, we 
notice that the size distribution of void filaments becomes stabilized 
when the pruning level is $p\ge 5$. The origin of this difference in the 
value of $p$ lies in the fact that the galaxies correspond to the 
smaller-size substructures of the FOF halos. When the galaxies are used 
to construct the MSTs, the MSTs have more twigs and more superfluous 
small-scale noise. That is why a higher level $p=5$ is required to 
prune the MSTs.

\begin{figure}
\begin{center}
\includegraphics[width=84mm]{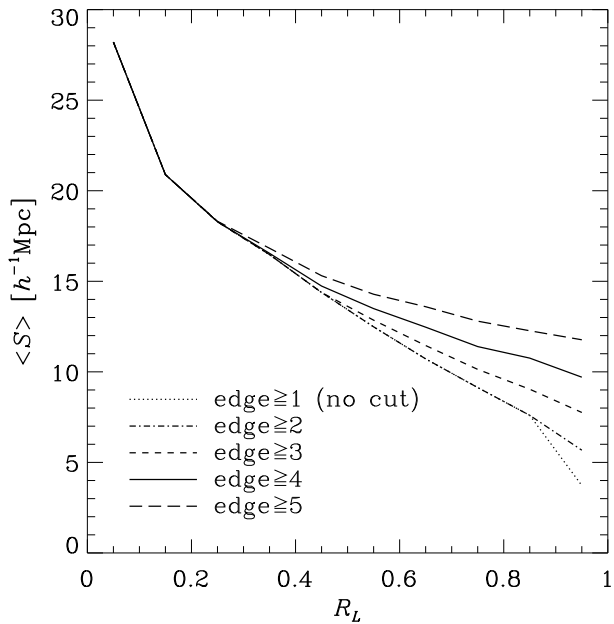}
\caption{Mean sizes of the void filaments identified in the Millennium Run
semi-analytic galaxy catalogue as a function of the filament's linearity
$R_{L}$ for various edge-thresholds.}
\label{fig:size}
\end{center}
\end{figure}

The size of each filament, $S$, is measured as the spatial extent of the 
filament galaxies (see eq.[1] in PL09), and the linearity of a given filament 
$R_{L}$, which is introduced to quantify the filament straightness, is 
measured as the ratio of the end-to-end distance to the total length
\citep{bar-etal85}. The range of $R_{L}$ is $[0,1]$ and the degree
of the filament straightness increases with $R_{L}$.
Note that the value of $R_{L}$ is biased toward the small value of
$S$. The shortest filaments that contain only one edge (i.e., two
nodes) have $R_{L}=1$ by definition. To study the correlations
between the properties of void galaxies and the linearity of void
filaments, it should be required to select only long filaments since
the short filaments are dominant in the high linearity section,
causing a bias in the measurement of correlations. In other words,
to find true correlations between $R_{L}$ and void galaxy's
properties, it is necessary to put some lower limit ($n_{c}$) on the
number of filament's edges. Fig.~\ref{fig:size} plots the mean sizes
of the void filaments as a function of $R_{L}$, for the five
different cases of the edge-threshold ($n_{c}=1,\ 2,\ 3,\ 4$ and $5$
as dotted, dot-dashed, dashed, solid and long dashed line
respectively). As can be seen, for all cases $\langle S\rangle$
decreases as $R_{L}$ increases. But, for $n_{c}\ge 4$ (long-dashed
and solid lines), $\langle S\rangle$ does not decreases
significantly with $R_{L}$ in the high-linearity range of $R_{L}\ge
0.7$. Using this empirical result, we choose $n_{c}=4$ and select
only those filaments which consist of $4$ edges or more (i.e., five
galaxies or more).

\begin{figure*}
\begin{center}
\includegraphics[width=175mm]{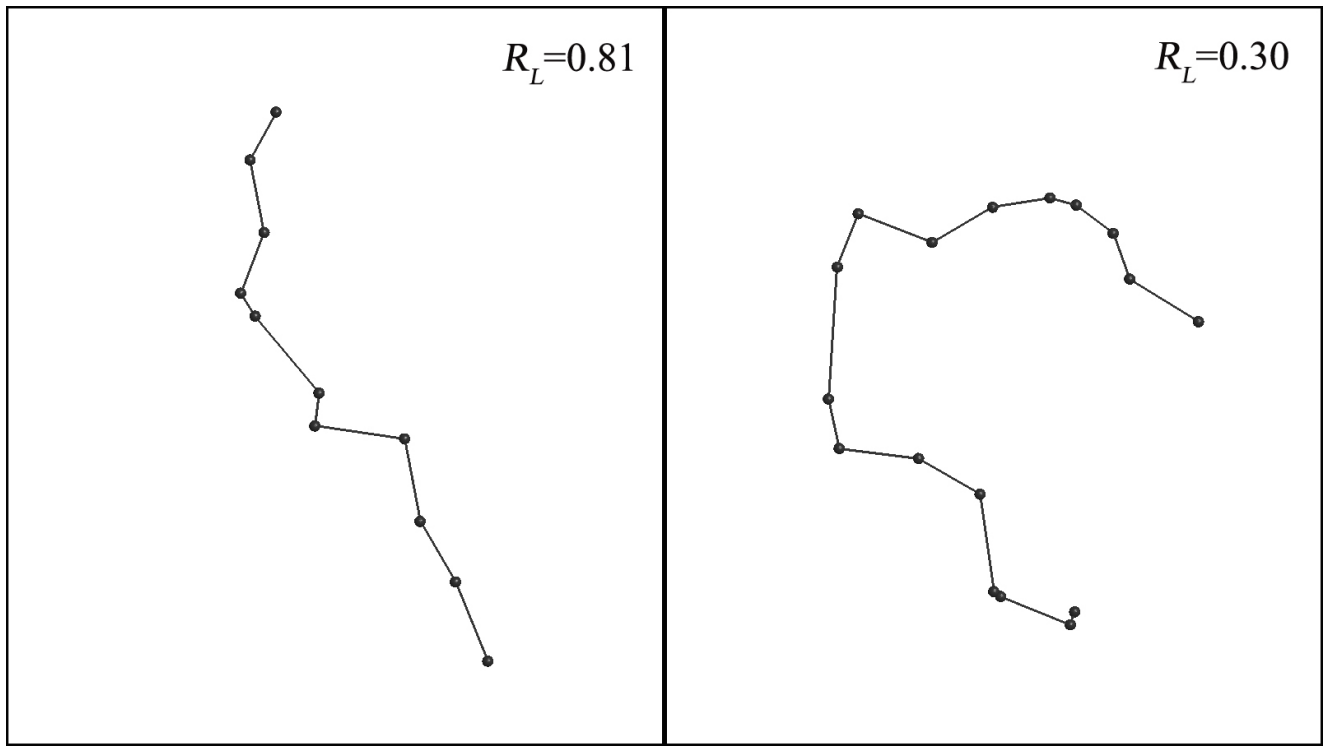}
\caption{Illustration of the two examples of void filaments having
two different values of the filament linearity, $R_{L}$.}
\label{fig:ill}
\end{center}
\end{figure*}

Table~\ref{tab:fil} lists the statistical properties of the void
filaments. The first row corresponds to the case where no edge
threshold is applied (one edge or more), while the second one
corresponds to the case that only those filaments with $n_{c}=4$ 
are considered. As can be seen, the mean linearity of the void
filaments is low when the edge-threshold, $n_{c}=4$, is applied. It
indicates that the short-size filaments dominate the high-linearity
section. Fig.~\ref{fig:ill} illustrates two examples of the void
filaments with two different values of $R_{L}$, demonstrating
clearly that the higher the value of $R_{L}$ is, the more straight a
filament is.

\begin{table}
\centering \caption{Edge-cut $n_{c}$, total number of void filaments $N_{f}$,
mean number of void filaments per a void ($\bar{n}_{f}$), mean length
of void filaments ($\bar{R}_{f}$) in unit of $h^{-1}$Mpc, and mean
linearity of void filaments ($\bar{R}_{L}$). }
\begin{tabular}{@{}ccccc}
\hline
$n_{c}$ & $N_{f}$ & $\bar{n}_{f}$ & $\bar{R}_{f}$ & $\bar{R}_{L}$\\
 & & & [$h^{-1}$Mpc] & \\
\hline
1 & $58554$ & $4$ & $9.2$ & $0.73$\\
4 & $29755$ & $2$ & $13.8$ & $0.56$\\
\hline
\end{tabular}
\label{tab:fil}
\end{table}

\section{THE BRIDGE EFFECT}

\subsection{Linearity vs. Galaxy Property}

The void filaments correspond to deep potential wells within which
excess of matter and gas can exist in otherwise extremely underdense
environments. They bridge the void galaxies onto which matter and gas
from the surrounding large scale structures accrete and flow along
their longest-axis directions.  Therefore, the coherent accretion
and flow of matter and gas may occur more efficiently in more straight
filaments. It is expected that there may be a link between the physical
properties of void galaxies and the linearity of their parent filaments,
caused by this bridge effect of void filaments.

\begin{figure*}
\begin{center}
\includegraphics[width=175mm]{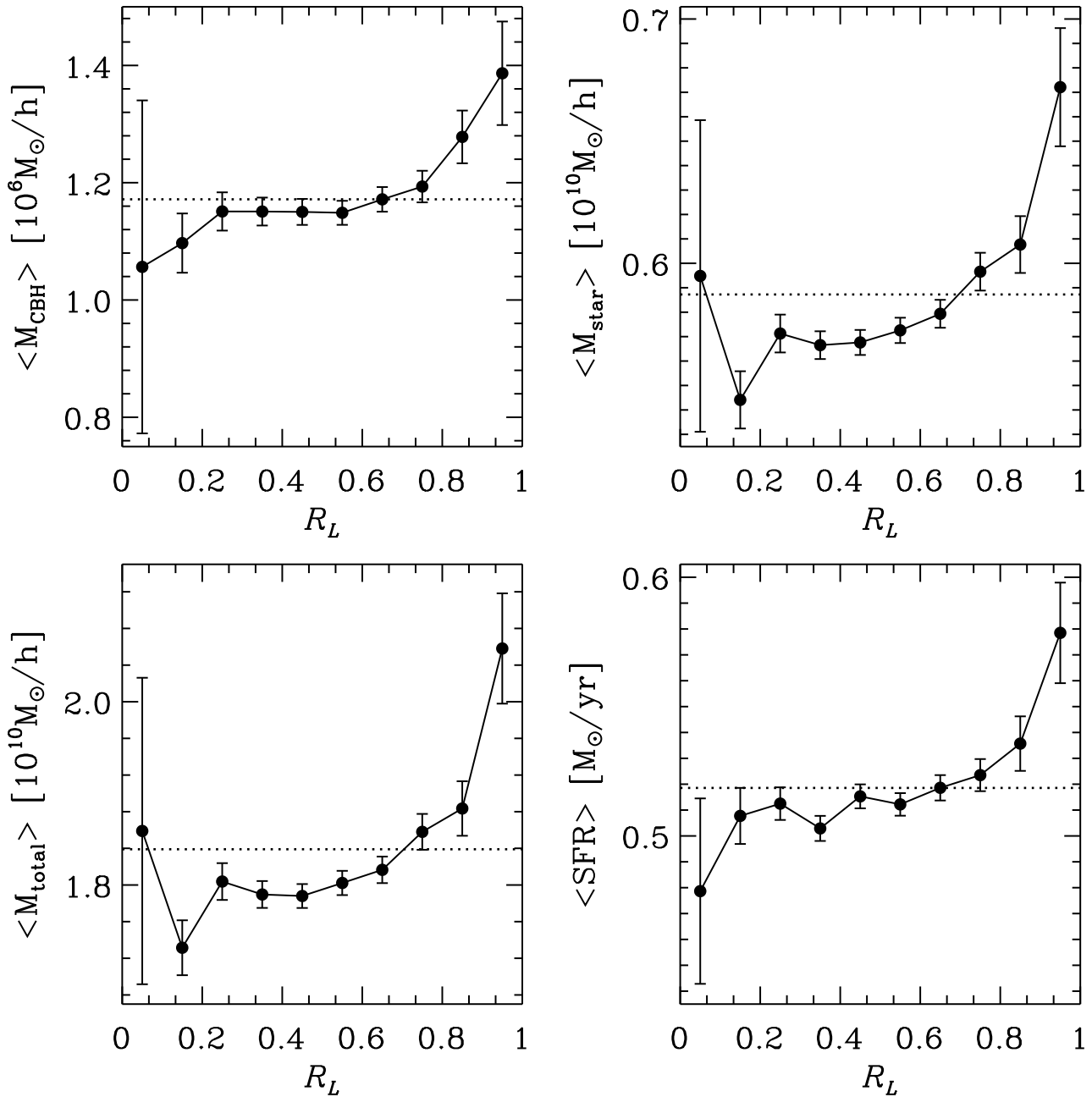}
\caption{Means of the central black hole mass (top-left), the stellar mass
(top-right), the total mass (bottom-left), and the star formation rate
(bottom-right) averaged over the constituent galaxies as a function of the
host filament's linearity. In each panel the horizontal dotted line
corresponds to the mean values averaged over all filaments.}
\label{fig:cor}
\end{center}
\end{figure*}

We measure $R_L$ of the selected void filaments and bin the values
of their linearity $R_{L}$. Then, we calculate the mean values of
central black hole mass ($M_{\rm CBH}$), stellar mass ($M_{\rm
star}$), total mass ($M_{\rm total}$), and the star formation rate
(${\rm SFR}$) of the void galaxies constituting a given void
filaments with $R_{L}$ in a differential range of
$[R_{L},R_{L}+dR_{L}]$. Here $M_{\rm total}$\footnote{$M_{\rm total}
= M_{\rm star} + M_{\rm bulge} + M_{\rm coldGas} + M_{\rm hotGas} +
M_{\rm ejected} + M_{\rm CBH}$} means the sum of all gas masses in the
catalogue. Fig. \ref{fig:cor} plots the means of the central black
hole mass (top-left), stellar mass (top-right), total gas mass
(bottom-left), and star formation rate (bottom-right) as a function
of parent filament's linearity, $R_{L}$. In each panel the dotted
line corresponds to the mean value averaged over all void filaments.
The errors are calculated as $\langle\Delta
P^{2}_{G}\rangle^{1/2}/\sqrt{N_{b}}$ where $N_{b}$ is the number of
void filaments belonging to a given bin of $R_{L}$ and $P_{G}$
represents a given galaxy property.  As can be seen, the four
different properties show a similar trend. The mean value of
galaxy's property increases as the value of $R_{L}$ increases. That
is, the void galaxies constituting more straight filaments tend to
be more luminous, more massive, having more massive central black
holes and higher star formation rate.

To see which property is most strongly correlated with $R_{L}$, we calculate
the linear correlation coefficient between $R_{L}$ and $P_{G}$ as:
\begin{equation}
\label{eqn:xi}
\xi_{P_G}\equiv\frac{\langle R_{L}P_{G}\rangle}
{\sqrt{\langle\Delta R^{2}_{L}\rangle\langle\Delta P^{2}_{G}\rangle}}.
\end{equation}
Table \ref{tab:linco} lists the values of $\xi_{P_G}$ for the four
different galaxy properties. As can be seen, the central black hole
mass is most strongly correlated with $R_{L}$. The star formation
rate is also quite strongly correlated with $R_{L}$. This results
indicate that the the coherent matter-accretion and gas-flow along
the straight filaments feed the central black holes and trigger the
star formation efficiently.
\begin{table}
\centering \caption{The linear correlation coefficients.}
\begin{tabular}{@{}cccc}
\hline
$\xi_{\rm CBH}$ & $\xi_{\rm star}$ & $\xi_{\rm total}$ & $\xi_{\rm SFR}$\\
\hline
$0.89$ & $0.69$ & $0.67$ & $0.85$\\
\hline
\end{tabular}
\label{tab:linco}
\end{table}

\subsection{Linearity vs. Halo Mass}

Now that strong correlations between the linearity of void filaments
and the physical properties of void galaxies are found, we would like
to investigate the dependence of the mass of void halos on the linearity
of their filament's linearity, using the catalogue of the filaments of
void halos that was constructed in PL09.
For each filament of void halos, we determine the maximum mass ($M_{max}$)
and the minimum mass ($M_{min}$) among the masses of the constituent halos.
Then, we calculate the means of $M_{max}$ and $M_{min}$ as a function of
$R_{L}$.

\begin{figure*}
\begin{center}
\includegraphics[width=175mm]{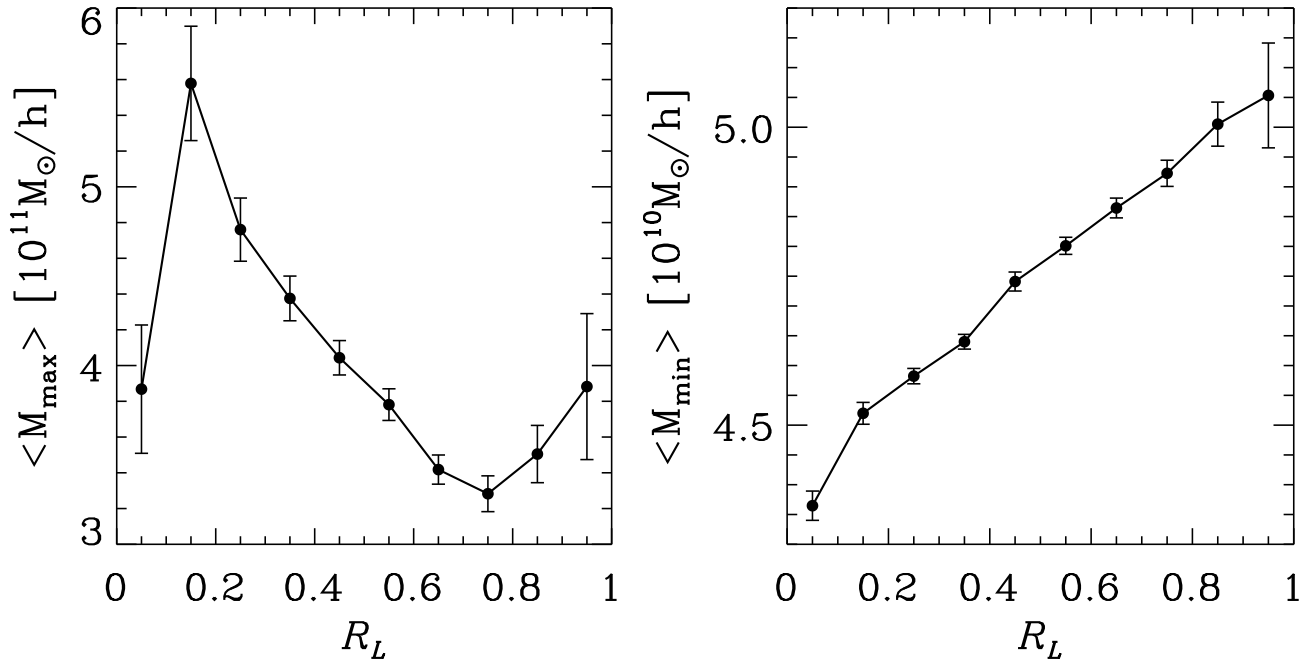}
\caption{Means of the maximum (left) and minimum (right) masses of
void halos as a function of the linearity of their parent's filaments,
$R_{L}$.}
\label{fig:maxmin}
\end{center}
\end{figure*}

\begin{figure}
\begin{center}
\includegraphics[width=84mm]{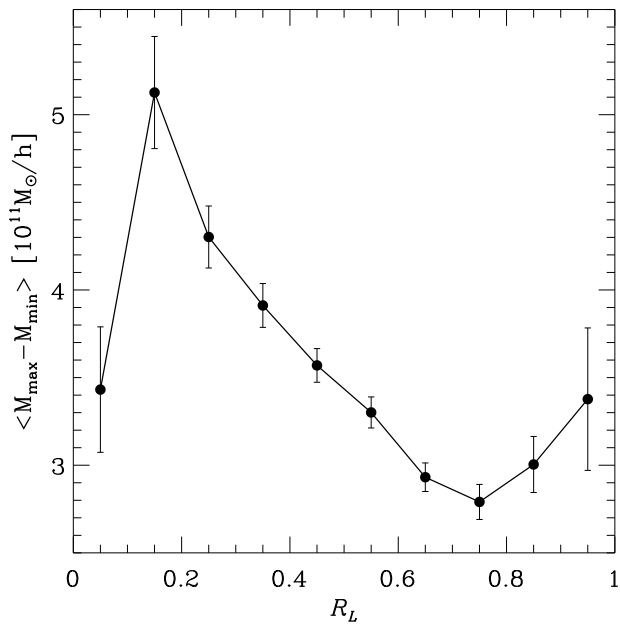}
\caption{Means of the differences between the maximum and minimum masses of
void halos as a function of the linearity of their parent's filaments,
$R_{L}$.}
\label{fig:dif}
\end{center}
\end{figure}

Fig.~\ref{fig:maxmin} plots $\langle M_{max}\rangle$ and
$\langle M_{min}\rangle$ as a function of $R_{L}$ in the left and
right panels, respectively. The errors in each panel is calculated
as $\langle\Delta M^{2}\rangle^{1/2}/\sqrt{N_{b}}$.
As can be seen, the value of $\langle M_{max}\rangle$ decreases almost
monotonically with $R_{L}$, while the value of $\langle M_{min}\rangle$
increases monotonically with $R_{L}$. It indicates that the differences
among the masses of void halos are smaller when the parent filaments are
more straight. To see this more clearly, we calculate the mean values of
$M_{max}-M_{min}$ as a function of $R_{L}$ which is plotted in
Fig.~\ref{fig:dif}. It is now obvious that the difference decreases
as $R_{L}$ increases. In other words, the dark halos that constitute
straight filaments tend to have similar masses. The efficient
matter-accretion along straight filaments tend to synchronize the masses
of void halos and regulate the shapes of potential wells of the void
filaments to be flat.

\section{SUMMARY AND DISCUSSION}

We have found strong correlations of the stellar mass, central black hole
mass and star formation rate of void galaxies with the linearity of their
parent filaments. Among the three properties of void galaxies, the masses
of the central black holes are mostly strongly correlated with the linearity
of void filaments. The existence of this cross-correlations indicates the
bridge effect of void filaments. Matter and gas from the surrounding
large-scale structures accrete and flow onto the void galaxies more
efficiently along the more straight filaments. The dark halos constituting
straight filaments are found to have similar masses, which implies that the
potential wells generated by straight void filaments have rather flat-shaped
bottoms. The efficient accretion of matter along straight filaments tend to
reduce mass difference among the constituent void halos.

It is interesting to note that our results may provide an explanation on
high specific star formation rate of void galaxies and high activity of void 
AGNs found in data from the Sloan Digital Sky Survey \citep{con-etal08,
roj-etal05}. The void filaments should be more straight on average than the 
wall filaments since the nonlinear effect that can distort the filaments is 
less strong in void regions. Thus, matter-accretion and gas-flow along more
straight void filaments are likely to supply more efficiently fuel for the 
stars and AGNs of the void galaxies.

It is, however, worth noting that our results are subject to the specific 
choice of the void-finding algorithm and the filament-finding algorithm. 
Unlike bound halos, neither voids nor filaments can be uniquely defined. 
As shown in the current seminal work of \citet{col-etal08}, the voids 
identified by the HV02 algorithm tend to be relatively large in size compared 
with those found by different void-finders. Thus, if other algorithms were
used, then it would have been difficult to find long filaments in
voids. 

Besides, our results are based on numerical data from the Millennium Run 
simulations. To confirm the existence of the bridge effect of void filaments, 
an observational analysis dealing with real data is required. Unfortunately, 
however, it is quite unlikely to obtain any meaningful statistics about the 
bridge effect of void filaments from the currently available observational 
data. As first pointed out by \citet{pee01}, the observed voids look 
apparently much emptier than the simulated voids, containing very few 
galaxies \citep[e.g.,][]{FN09}. It is believed that this discrepancy 
between simulation and observation on voids indicates not the failure of 
the concordance cosmology but the limitation of current observational 
technique. Anyway it would be extremely difficult to find as many void 
filaments as necessary for statistical analysis from the current available 
observational data. Hence, we think that an observational test of our 
results is beyond the scope of this paper.
It is concluded that our numerical finding of the bridge effect of void 
filaments may provide a new insight about how galaxies form and evolve 
in a $\Lambda$CDM cosmology.

\section*{Acknowledgments}

We thank an anonymous referee for helpful comments.
The Millennium Run simulation used in this paper was carried out by the
Virgo Supercomputing Consortium at the Computing Center of the Max-Planck
Society in Garching. The Millennium Simulation data are available at
http://www.mpa-garching.mpg.de/millennium.
This work is financially supported by the Korea Science and Engineering
Foundation (KOSEF) grant funded by the Korean Government
(MOST, NO. R01-2007-000-10246-0).



\begin{thebibliography}{dummy}

\bibitem[Barrow et al.(1985)]{bar-etal85}
Barrow, J.~D., Bhavsar, S.~P., \& Sonoda, D.~H., 1985, MNRAS, 216, 17
\bibitem[Bastian et al.(2007)]{bas-etal07}
Bastian, N., Ercolano, B., Gieles, M., Rosolowsky, E., Scheepmaker,
R.~A., Gutermuth, R., \& Efremov, Yu., 2007, MNRAS, 379, 1302
\bibitem[Bhavsar \& Ling(1988a)]{bha-lin88a}
Bhavsar, S.~P., \& Ling, E.~N., 1988a, ApJl, 331, 63
\bibitem[Bhavsar \& Ling(1988b)]{bha-lin88b}
Bhavsar, S.~P., \& Ling, E.~N., 1988b, PASP, 100, 1314
\bibitem[Bond et al.(1996)]{bon-etal96}
Bond, J., R., Kofman, L., \& Pogosyan, D., 1996, Nature, 380, 603
\bibitem[Colberg et al.(2008)]{col-etal08}
Colberg, J.~M. et al. 2008, MNRAS, 387, 933
\bibitem[Coles et al.(1998)]{col-etal98}
Coles, P., Pearson, R.~C., Borgani, S., Plionis, M., \& Moscardini,
L., 1998, MNRAS, 294, 245
\bibitem[Colless et al.(2001)]{col-etal01}
Colless et al., 2001, MNRAS, 328, 1039
\bibitem[Constantin et al.(2008)]{con-etal08}
Constantin, A., Hoyle, F., \& Vogeley, M.~S., 2008, ApJ, 673, 715
\bibitem[Desjacques(2008)]{des08}
Desjacques, V.\ 2008, MNRAS, 388, 638
\bibitem[D'Elia et al.(2008)]{del-etal08}
D'Elia, V., Fiore, F., Mathur, S., \& Cocchia, F., 2008, A\& A, 484, 303
\bibitem[Einasto(2006)]{ein06}
Einasto, J., 2006, Formation of the Supercluster-Void Networks,
proceedings of the Detre Centennial Conference : Communications from
the Konkoly Observatory, ed. L.-G. Balazs, L. Szabados, A. Holl
(CoKon Homepage), 104
\bibitem[El-Ad \& Piran(1997)]{ela-pir97}
El-Ad, H., \& Piran, T., 1997, ApJ, 491, 421
\bibitem[Foster \& Nelson(2009)]{FN09} 
Foster, C., \& Nelson, L.~A.\ 2009, ApJ, 699, 1252 
\bibitem[Graham et al.(1995)]{gra-etal95}
Graham, M.~J., Clowes, R.~G., \& Campusano, L.~E., 1995, MNRAS, 275, 790
\bibitem[Gao et al.(2005)]{gao-etal05}
Gao, L., Springel, V., \& White, S.~D.~M., 2005, MNRAS, 363, L66
\bibitem[Hahn et al.(2008)]{hah-etal08}
Hahn, O., Porciani, C., Dekel, A., \& Carollo, C.~M., 2008, (arXiv:0803.4211)
\bibitem[Hoyle \& Vogeley(2002)]{hoy-vog02}
Hoyle, F., \& Vogeley, M.~S., 2002, ApJ, 566, 641
\bibitem[Krzewina \& Saslaw(1996)]{krz-sas96}
Krzewina, L.~G., \& Saslaw, W.~C., 1996, MNRAS, 278, 869
\bibitem[Lee \& Erdogdu(2007)]{lee-erd07}
Lee, J., \& Erdogdu, P., 2007, ApJ, 671, 1248
\bibitem[Lee \& Lee(2008)]{lee-lee08}
Lee, J., \& Lee, B., 2008, ApJ, 688, 78
\bibitem[Lee \& Park(2006)]{lee-par06}
Lee, J., \& Park, D., 2006, ApJ, 652, 1
\bibitem[Pandey \& Bharadwaj(2008)]{pan-bha08}
Pandey, B., \& Bharadwaj, S., 2008, MNRAS, 387, 767
\bibitem[Park \& Lee(2007a)]{par-lee07a}
Park, D., \& Lee, J., 2007, Phys. Rev. Lett., 98, 081301
\bibitem[Park \& Lee(2007b)]{par-lee07b}
Park, D., \& Lee, J., 2007, ApJ, 665, 96
\bibitem[Park \& Lee(2009)]{par-lee09}
Park, D., \& Lee, J., 2009, MNRAS,in press (arXiv.org:)
\bibitem[Pearson \& Coles(1995)]{pea-col95}
Pearson, R.~C., \& Coles, P., 1995, MNRAS, 272, 231
\bibitem[Peebles(2001)]{pee01}
Peebles, P.~J.~E., 2001, ApJ, 557, 495
\bibitem[Platen et al.(2008)]{pla-etal08}
Platen, E., van de Weygaert, R., \& Jones, B.~J.~T., 2008, MNRAS,387, 128
\bibitem[Plionis et al.(1992)]{pli-etal92}
Plionis, M., Valdarnini, R., \& Jing, Y.~P., 1992, ApJ, 398, 12
\bibitem[Porter et al.(2008)]{por-etal08}
Porter, S.~C., Raychaudhury, S., Pimbblet, K.~A.,
\& Drinkwater, M.~J., 2008, MNRAS, 388, 1152
\bibitem[Roberts et al.(2007)]{rob-etal07}
Roberts, S., Davies, J., Sabatini, S., Auld, R., \& Smith, R., 2007,
MNRAS, 379, 1053
\bibitem[Rojas et al.(2005)]{roj-etal05}
Rojas, R.~R., Vogeley, M.~S., Hoyle, F., \& Brinkmann, J.,
2005, ApJ, 624, 571
\bibitem[Sahni et al.(1994)]{sah-etal94}
Sahni, V., Sathyaprakash, B.~S., \& Shandarin, S.~F., 1994, ApJ, 431, 20
\bibitem[Sahni \& Shandarin(1996)]{sah-sha96}
Sahni, V., \& Shandarin, S.~F., 1996, MNRAS, 282, 641
\bibitem[Shandarin et al.(2004)]{sha-etal04}
Shandarin, S.~F., Sheth, J.~V., \& Sahni, V., 2004, MNRAS, 353, 162
\bibitem[Shandarin et al.(2006)]{sha-etal06}
Shandarin, S., Feldman, H.~A., Heitmann, K., \& Habib, S., 2006,
MNRAS, 367, 1629
\bibitem[Springel et al.(2005)]{spr-etal05}
Springel, V. et al., 2005, Nature , 435, 629
\bibitem[Tanaka et al.(2007)]{tan-etal07}
Tanaka, M., Hoshi, T., Kodama, T., \& Kashikawa, N., 2007, MNRAS, 379, 1546

\label{lastpage}
\end{thebibliography}
\end{document}